\documentclass{svproc}
\usepackage{url}

\usepackage{algorithm2e}
\usepackage{amsmath}
\usepackage{multirow}
\usepackage[table,xcdraw]{xcolor}
\usepackage{graphicx}

\usepackage{hyperref}

\usepackage{tcolorbox}

\begin{document}
\mainmatter              % start of a contribution
\title{Pattern Analysis of Money Flows\\in the Bitcoin Blockchain}
\titlerunning{Bitcoin Money Flows}  % abbreviated title (for running head)
\author{Natkamon Tovanich\inst{1} \and Rémy Cazabet\inst{2}}
\authorrunning{Tovanich and Cazabet} % abbreviated author list (for running head)
\tocauthor{Natkamon Tovanich and Rémy Cazabet}
\institute{Blockchain \& B2B Platforms Chair, École Polytechnique,\\ Institut Polytechnique de Paris, 91120 Palaiseau, France, \email{natkamon.tov@gmail.com},\\
\and
Univ de Lyon, CNRS, Université Lyon 1, LIRIS, UMR5205,\\69622 Villeurbanne, France, \email{remy.cazabet@gmail.com}}

\maketitle              % typeset the title of the contribution

\begin{abstract}
Bitcoin is the first and highest valued cryptocurrency that stores transactions in a publicly distributed ledger called the blockchain. Understanding the activity and behavior of Bitcoin actors is a crucial research topic as they are pseudonymous in the transaction network. In this article, we propose a method based on taint analysis to extract taint flows---dynamic networks representing the sequence of Bitcoins transferred from an initial source to other actors until dissolution. Then, we apply graph embedding methods to characterize taint flows. We evaluate our embedding method with taint flows from top mining pools and show that it can classify mining pools with high accuracy. We also found that taint flows from the same period show high similarity. Our work proves that tracing the money flows can be a promising approach to classifying source actors and characterizing different money flow patterns.
\end{abstract}

\section{Introduction}

Bitcoin is the oldest and most used cryptocurrency, attracting broad interest from the general public and researchers. In contrast to traditional financial networks, transactions can be observed by anyone on the public blockchain, on which users exchange Bitcoins pseudonymously.
This data allows researchers to study economic activities in fine detail.
One of the objectives of those research is to understand how the Bitcoin socio-technical system works, particularly, 1) Who are the important actors of the Bitcoin economy? \cite{lischke2016analyzing,liu2021characterizing,meiklejohn2013fistful}; 2) How is the network of transactions organized? \cite{lischke2016analyzing,nerurkar2021dissecting,vallarano2020bitcoin}; and 3) How to identify and track illegal activity? \cite{chainalysis2022,bartoletti2021cryptocurrency,weber2019anti}.

Tracing the flow of money---where the money goes, to whom, and when---is also an essential task in cryptocurrencies and critical for financial forensics to trace money from suspicious sources and characterize different users' behaviors.
%\nat{However, this approach is still understudied to deanonymize and characterize actors from unknown sources.} %However, it is still an understudied and challenging question in the Bitcoin research domain despite pseudonymous actors. 
We investigate two main questions regarding the relationship between the money source and subsequent transactions: 1) Does money flow differently in the Bitcoin network depending on its source?; and consequently, 2) Can we characterize a source actor given the observation of the flow of its coins in the network?

We propose an original way to synthesize the money flow from a given source into a concise dynamic network called a \textit{taint network}. We subsequently apply whole graph embedding methods to automatically assign taint networks to their origin actor. Beyond the demonstration that each actor has a characteristic flow allowing us to recognize its position in the network \cite{ahmed2018tendrils,meiklejohn2013fistful,tironsakkul2019probing}, our method can also be helpful for actor tracking as well as actor deanonymization tasks. The embedding of money flows from different actors is a promising feature for downstream tasks in machine learning models to classify the role of actors \cite{jourdan2018characterizing,zola2019cascading} or predict illegal transaction activities in the Bitcoin blockchain \cite{weber2019anti}.%bartoletti2021cryptocurrency

\section{Related Work}
\label{relatedWork}

Due to the pseudonymity of Bitcoin actors, the main research challenges concentrate on 1) deanonymizing those actors and 2) characterizing their roles in the transaction network. Early works proposed clustering heuristics to deanonymize addresses likely to belong to the same actors \cite{cazabet2017tracking,ermilov2017automatic,reid2013analysis,zhang2020heuristic}. 
%In reality, actors can use mixing services to hide traces that we need to filter out \cite{goldfeder2017cookie,wu2021detecting,wu2021towards}.
%Early works propose clustering heuristics to group addresses likely to hold by the same actor based on the transaction network property. The input address heuristics assume that input addresses of the same transaction belong to the same actor, which has been proven effective and widely adopted as the standard method to identify actors. Nonetheless, the mixing transactions need to be filtered out to avoid mistakenly grouping addresses that use this kind of service to hide their traces. In this work, we use input address heuristics with CoinJoin filtering implemented in BlockSci to identify actors.

Our work focuses on identifying and characterizing actors in the transaction network.
Previous works applied graph analysis and machine learning to classify the role of actors or whether the transactions are illicit or not.
Most works derived a set of descriptive features (e.g., the number and frequency of transactions, in- and out- degrees, and the number of different addresses used) \cite{akcora2019bitcoinheist,bartoletti2018data,harlev2018breaking,liu2021characterizing,michalski2020revealing} or the high order moments of transaction time \cite{lin2019evaluation}. These approaches rely on the actor's behavior, which easily manipulates it to hide its activities.

Other approaches thus rely on graph motifs, i.e., the set of subgraph patterns describing the neighbors of an actor  \cite{jourdan2018characterizing,ranshous2017exchange,wu2021detecting,zola2019cascading}. Due to computational reasons, those works construct the static graph features only from direct neighbors (2-motif) or neighbors of their neighbors (3-motif). They do not use the identity of these neighbors but simply numeric descriptions (e.g., the total amount sent or received and transaction fees). Node2vec has been used to embed the actor position in the address network \cite{michalski2020revealing}. Nonetheless, a few works include the temporal aspect to impose a temporal locality constraint on the motifs \cite{weber2019anti,wu2021detecting}.

Contrary to these approaches, our proposed method does not rely on the actor's activity or its direct neighbors but on the temporal network describing the whole flow of coins sent by an actor.
%Our method considers an actor during its entire existence---a problematic approach because bitcoin properties change radically along time due to evolution price, popularity, mining fees, etc.
%Moreover, the actors cannot control the flow directly to alter the result after transferring the coins to other actors.
We rely on the principle of tainted flows to trace the coins from source actors. Taint analysis has been used most notably in the context of tracking money from illegal sources \cite{ahmed2018tendrils,balthasar2017analysis,di2015bitconeview,moser2013inquiry,tironsakkul2019probing}. 
However, those works mainly focus on the destination of tainted coins. Our work expands this approach to analyze the full money flow from multiple sources. We are not merely interested in the destination of tainted coins but in characterizing the temporal networks created by those flows.

\section{\textit{Taint Flow} extraction}
\label{taintFlow}

Our objective in this work is to design a new method to characterize a \textbf{bitcoin source} based on the flow of its coins in the transaction network. The underlying hypothesis is that the way a coin travels in the transaction network is characteristic of its source. 
However, actors reached by tainted coins after an indefinite time and coins diffused in a more considerable amount of bitcoins in a transaction cannot be characteristics of their origin.
After presenting the tainting process on Bitcoin's blockchain, we thus introduce two principles to keep only relevant information in the flow: \textit{dissolved} coins and \textit{tag actors}.

\subsection{Bitcoin taint flow}
Bitcoin uses the unspent transaction output (UTXO) transaction model \cite{nakamoto2009bitcoin}. According to this model, transactions do not transfer money from one account to another. Instead, each output---each UTXO---of the transaction represents an amount of coin belonging to a known bitcoin address---a cryptographic public key. The rightful owner of the UTXO uses the corresponding private key to claim the money. It spends the UTXO(s) by signing them as input(s) in a new transaction and sends new UTXO output(s) to recipients' addresses.

A Bitcoin \emph{transaction network} can be modeled as a chain of UTXOs. A transaction ($tx$) is represented as a node. A directed edge represents a transfer of UTXO(s) from one transaction to another. We will refer to the in-edge and out-edge of $tx$ as input ($tx.in$) and output ($tx.out$), respectively. Each UTXO edge ($e$) is characterized by the amount of Bitcoin ($e.amount$) and the owner's address of that UTXO ($e.address$). It also contains references to the receiving ($e.receive$) and spending transaction nodes ($e.spend$).

%We define a \textbf{taint flow} as the \textbf{directed acyclic graph (DAG)} describing coin flow from a \textbf{source of interest} until \textbf{dissolution}.
We define the \emph{taint flow} as a directed acyclic graph ($G_{flow}$) tracing the sequence of transactions from a \textbf{source of interest} until \textbf{dissolution}. The source of interest can be one or several actors and limited to a given time interval based on the focus of the study. To construct a money flow, we recursively taint all UTXO outputs ($tx.out$) until the coins are \textbf{dissolved}.
%defined as a set of transactions ($\tau_o = \{tx_1, tx_2, ..., tx_n\}$) 

\begin{definition}[Dissolution]
We consider that a tainted coin is \textbf{dissolved} when its future positions in the transaction network will no longer be characteristic of its original position. More formally, a coin is dissolved when it is spent in a transaction with a purity value below a minimum threshold.
\end{definition}

\emph{Purity} measure ($\rho$) has been used to determine when the money is dissolved and stop following the transaction outputs \cite{di2015bitconeview}. Purity is the percentage of tainted money from the origin transaction set, defined as:

\begin{equation}
\rho(tx) = \frac{\sum_{e\in tx.in} \rho(e.receive) \cdot e.value}{\sum_{e\in tx.in} e.value}
\end{equation}
\label{purity}

%In the following, we use a purity threshold $\rho_{min} < 0.001$
The purity of a transaction without inputs is 1 by definition because it is the root transaction in the transaction flow. 
In this study, we set a purity threshold $\rho_{min} = 0.001$, which means that a coin is considered dissolved when it is spent in a transaction together with 1,000 times the amount of un-tainted coins. 
Besides, we stop following the flow when the transaction is $>$ 1 year apart from the source transactions ($time_{max}$). \autoref{alg:payout_flow} describes the process of retrieving transaction outputs and adding them to the money flow graph.

\begin{algorithm}
\SetAlgoLined
\SetKwInOut{Input}{Input}
\SetKwInOut{Output}{Output}
\Input{$\tau_o$ is a payout transaction as a seeding node of the payout flow.}
\Input{$\rho_{min}$ is a minimum purity threshold.}
\Input{$time_{max}$ is a maximum time threshold.}
\Output{$edges$ is the edge list of the payout flow.}
$queue\gets PriorityQueue([\tau_o])$\;
$edges\gets List()$\;
\While{$queue$ is not empty}{
    $tx\gets queue.pop()$\;
    \If{$\rho(tx) \geq \rho_{min}$ and $tx.time \leq time_{max}$}{
        \For{$e$ in $tx.out$}{
            $e.amount_{flow}\gets e.amount\times \rho(tx)$\;
            $edges.append(e)$\;
            $queue.append(e.spend)$\;
        }
    }
}
\caption{Reward payout flow extraction}
\label{alg:payout_flow}
\end{algorithm}

Our algorithm applies \textit{haircut tainting}, which assumes that the tainted money is divided equally to all output transactions in proportion to their amount \cite{ahmed2018tendrils,tironsakkul2019probing}. %The method maximizes tainted transactions to expand the $G_{flow}$. %We use the purity criteria and tag actors to prune the flow.

\subsection{Actors and Tag Actors}
\label{tag_actors}

%In practice, the owner can always change the address at negligible cost to hide their money trace. 
We defined actors as a set of addresses corresponding to a person, a group of persons, an organization, or any other entity owning a set of private keys to claim the ownership of UTXOs from public key addresses.
%The process of discovering those actors is a classic problem in the literature.
A simple but effective heuristic assumes that the input addresses in a transaction should belong to the same owner \cite{reid2013analysis,harrigan2016unreasonable}. We use the input address clustering heuristic implemented in the BlockSci library \cite{kalodner2020blocksci} that also filter CoinJoin transactions \cite{goldfeder2017cookie} to discover a set of addresses ($e.address$) belonging to the same actor (also called address cluster, $e.cluster$).
When analyzing a tainted flow, the relevant information is the actors involved in this flow. Therefore, a flow is summarized as a set of transactions between actors.

A taint flow can be large and sparse, making it difficult to compare with other flows \cite{ahmed2018tendrils}. To improve on this limit, we propose to work on variants of flows in which we keep only important actors named \textit{tag actors}.

 \begin{definition}[Tag Actor]
To characterize a flow, we can describe it using a subset of all encountered actors, called \textbf{tag actors}. Tag actors are prominent ones that are likely to stay constant in time and to be reached by many flows from different sources.
 \end{definition}

We propose two ways of defining tag actors: 1) \textbf{frequent actors} consist in keeping a fraction of the most frequent actors; and 2) \textbf{known actors} are chosen based on external data.
%They are actors found by the same heuristic as others but validated as corresponding to an identified known entity. They are more reliable than those found by the heuristic. It also allows using additional information, such as the type of actor (e.g., exchange platform, gambling service, marketplace, etc.), which is known from the external source.
In this work, we use WalletExplorer dataset \cite{walletexplorer} that provides a collection of 375 known actors, in particular services or companies, linking their addresses with the name and type of the service (e.g., exchange platform, gambling service, and marketplace).

\section{Taint Flow Embedding}
\label{embedding}

%Our objective is to assess the similarity of taint flows, such as two flows reaching similar actors in a similar manner are considered similar, and conversely, two flows reaching different actors and/or same actors in a different temporal way are considered dissimilar.
Since taint flows are represented as graphs, we use whole graph embedding approaches to assess the similarity of the taint flows and characterize their patterns.
The principle of those methods, such as Graph2vec \cite{narayanan2017graph2vec} or Anonymous walk embedding \cite{ivanov2018anonymous} is to assign a low-dimensional vector representation to each graph such as two graphs considered similar according to a chosen network structure representation are close in the resulting embedding space. 

%Graph embedding is a common approach to represent high-dimensional sparse graphs to lower-dimensional vector space \cite{xu2021understanding}. 
Recently, the Geo2DR methodology \cite{scherer2020learning} was introduced to allow one to design custom embedding methods and construct whole graph embedding. The methodology consists of two phases: 1) induction of descriptive substructure patterns and 2) learning of vector representations.
In this case, we deal with taint flows that have directed acyclic graphs, temporal nature, and different types of node labels. Therefore, we define our custom process to produce graph walks in the first phase.

\subsection{Induction of Descriptive Substructure Patterns}
\label{walk}
We use a random walk-based approach to extract substructure patterns from taint flows. We compare different variants of random walks and nodes labels vocabularies:
%\subsubsection{Walk types}
\begin{itemize}
    \item \textbf{RW} - Unbiased Random Walks. We generate random walks starting only from the source node, without considering weights, following edge temporal directions. A walk ends when encountering a dissolved node. 
    \item \textbf{SPW} - Shortest Path Walk. To generate an instance of the shortest path walk, we randomly choose a leaf node (dissolved) and walk through the shortest path between the source and a dissolved node.
\end{itemize}

\noindent We prune the walks based on the following sets of tag actors:

%\subsubsection{Pruning - Tag actors}
\begin{itemize}
    \item \textbf{All clusters}: We use no pruning and keep all actors.
    \item \textbf{Frequent clusters}: We keep only clusters that appear in more than 50\% of all taint flows as tag actors. The objective is to increase the fraction of shared vocabulary between flows to make learning more efficient. 
    \item \textbf{Known actors}: We use as tag actors only those known from an external source, WalletExplorer \cite{walletexplorer}. In the \textbf{name} variant, the node label corresponds to its name. In the \textbf{type} variant, the label corresponds to its type, one of exchange, wallet, service, marketplace, mixer, lending, and gambling.
\end{itemize}

We replace all mining pool clusters and names with a ``mining'' label to prevent the model from training the embeddings from the source actors.

\subsubsection{Temporal pattern}
In this variant, we use the same methods, but we integrate the time aspect into the walks, assuming that the time to reach different parts of the network might be characteristic of the source. We use a tuple (original label, time) where time is defined as $\lfloor \log_2(\Delta t)\rfloor$, with $\Delta t$ the time elapsed between the encoded transaction and the source transaction, in days. Examples of temporal pattern labels are thus: (ID: 63566, day: 7), (Name: Bitstamp.net, day: 7), or (Type: Exchange, day: 7). We use rounded log values to avoid sparse vocabulary.

\subsection{Learning Vector Representations}
%In the previous sections, we have introduced several variants of the substructure description extraction through walks in the graph. The second step of the process consists in learning the vector representation.
%For the second step, learning the vector representation from walks,
We use Distributed Memory Model of Paragraph Vectors (PV-DM) to train the embedding of the flow. PV-DM is one of the two variations of neural network models presented in the Doc2Vec paper \cite{le2014distributed}. The model is trained to maximize the prediction accuracy of the center vocable, given the surrounding vocables. We chose the PV-DM model because it preserves the order sequence of the walk rather than predicting a bag of words in a sentence in the PV-DBOW model. In our experiment, we set a typical embedding size ($n = 128$) to compare the different labeling strategies.

\section{Flow-based Actor Identification}
\label{actorIdentification}

In this section, we demonstrate how the taint flow embeddings can be leveraged to identify Bitcoin actors from transaction sources. We extract taint flows from Bitcoin mining pools and experiment with two actor-disambiguation tasks: 1) identification of source actors using supervised learning and 2) automatic discovery of actors based on clustering. Finally, we evaluate the capacity of the embedding to differentiate between temporal origins.

\subsection{Taint Flows of Bitcoin Mining Pools}
We focus our experiment on the identification of mining pools. Mining pools are among the most important actors in the Bitcoin ecosystem. They correspond to companies that regroup the activity of various miners---from individuals to mining farms---under a single entity, with the prospect of sharing the mining rewards obtained to mitigate the effect of chance on their source of revenue \cite{romiti2019deep,tovanich2022evolution}.
We chose mining pools because they are well-studied actors, persisting long enough in time, for which we can be confident in the data for validation. 

We extract taint flows from the top-3 mining pools for each month between 2013 and 2016 to represent different sources of flows from large representative mining pools. We chose this specific period because the WalletExplorer dataset stopped updating the actors used as known actors in 2016 \cite{walletexplorer}.

For each top-3 pool in a month, we taint all coins received from coinbase transactions---i.e., newly generated coins---on a random day of that month as a set of source transactions. We then construct taint flows and embed them with different methods according to the process defined in the previous sections.

As a result, our dataset consists of 144 taint flows from 11 mining pools. They are of various sizes and can be too large to use traditional embedding methods on standard computers. The average number of transactions is 611,327 (sd: 362,073, median: 569,576, max: 2,181,876) while the average number of clusters is 303,955 (sd: 178,145, median: 285,991, max: 1,131,919). There are 3,697 frequent clusters existing in more than $50\%$ of all flows.

\begin{figure}
\centerline{\includegraphics[width=\linewidth]{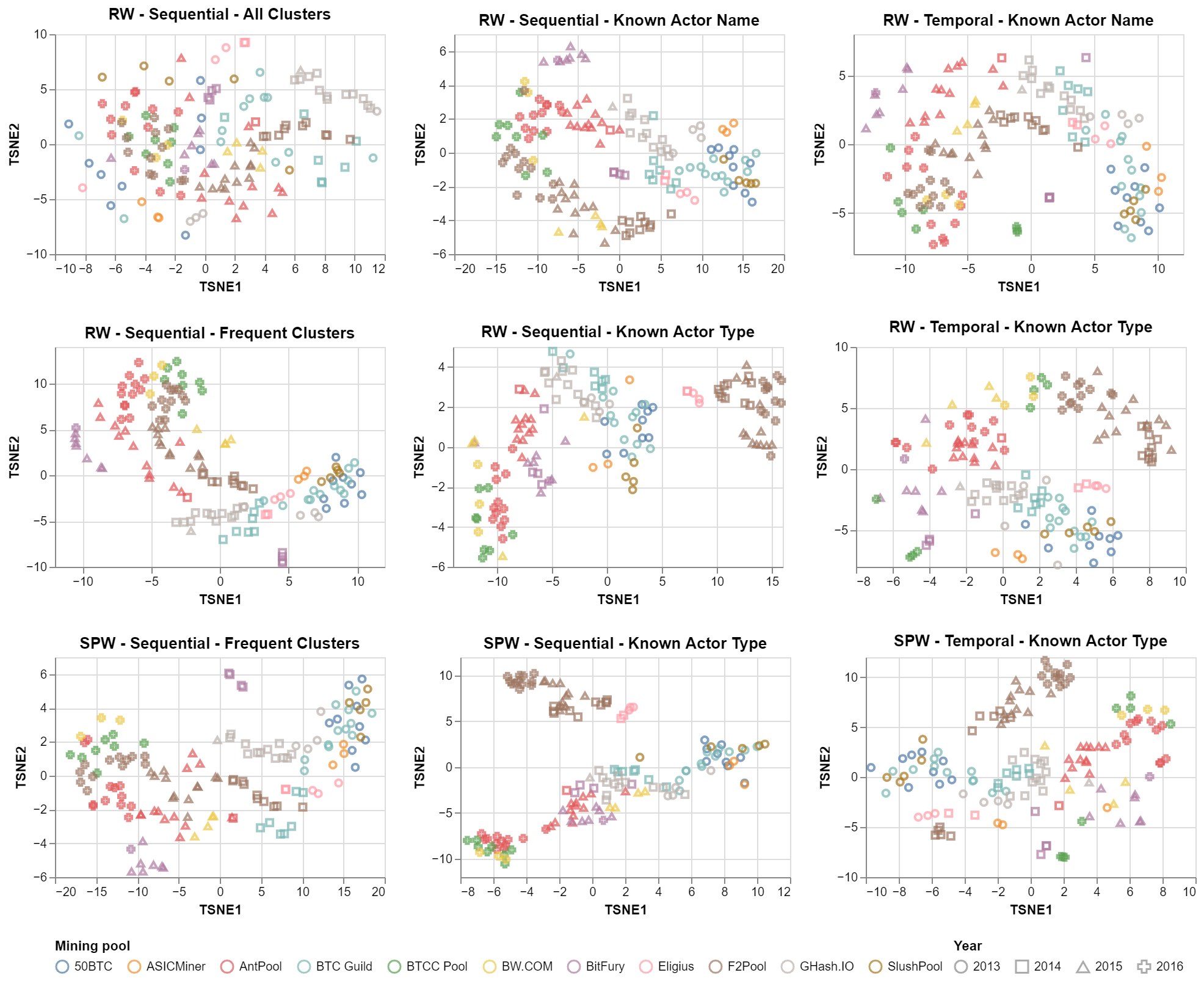}}
\caption{T-SNE projections of selected taint flow embeddings}
\label{fig:tsne}
\end{figure}

The result of the representation learning, embedded in two dimensions using t-SNE \cite{van2008visualizing}, is visualized in \autoref{fig:tsne}. In most cases, the multiple taint flows of the same mining pool (nodes of the same color) seem close in the embedding space. The method thus captures at least some elements of the identity of the source based on its taint flow. We also observe that the temporal aspect plays an important role and is well captured in the figures. For instance, in \textit{RW - Temporal - Known Actor Name} and \textit{RW - Sequential - Frequent Clusters} embedding, we see a shift---right to left---from circles to squares and triangles, and finally crosses, corresponding to the increasing years of the source.

%In the following sections, we quantify how actionable is this result for source recognition tasks.

% [Interpretation of the clustering result]

% To assess whether the embedding method can separate different taint flow patterns, we apply t-distributed stochastic neighbor embedding  to project embedding flows in 2-dimension scatterplot. \autoref{fig:random} display t-SNE projection of ...

%In particular, many pools tag their own transactions with their names, as a sort of \textit{advertisement} of their activity. We could rely on external sources such as WalletExplorer\footnote{\url{https://www.walletexplorer.com}} to work on different types of actors such as Exchanges or Gambling services. 
%Note that we focus only on newly generated coins from the source in this experiment, but the method applies without restriction on coins received from other actors.

\subsection{Actor Identification Task}

We train a k-Nearest Neighbor (k-NN) classification model with $k=3$ to identify mining pools from the embedding space. k-NN is a simple model that can capture non-linear decision boundaries. We have to keep $k$ small due to the relatively low number of observations.
%Without the complex classifier, we can infer how effectively the embedding method can help identify the source.
To evaluate the model performance, we use leave-one-out cross-validation (LOOCV), i.e., we consider that all sources are known but one and try to predict the identity of that unknown source from the others. %We repeat the prediction for each source.

\subsubsection{Baseline}
We define two baselines to compare the model performance with our embedding methods:

\begin{enumerate}
    \item \textbf{Actor network features:} We extract a set of descriptive features from all cluster networks, including the number of nodes and edges, density, and degree assortativity. We also calculate nodes in- and out- degrees, clustering coefficient, and eigenvector centrality and report the minimum, $1^{st}$--$3^{rd}$ quantiles, maximum, mean, standard deviation, and mean absolute deviation for each feature.
    \item \textbf{Graph2vec:} We train Graph2vec models \cite{narayanan2017graph2vec} to embed frequent clusters and known actor name networks. We cannot train the model with the \textit{All clusters} setting because the size of the graph can be enormous and make it impractical to compute the embedding.
\end{enumerate}

\begin{table}[]
\caption{Evaluation of taint flow embeddings of top-3 mining pools in 2013-2016}
\label{tab:evaluation}
\resizebox{\linewidth}{!}{
\begin{tabular}{l|c|c|c|c|c|c|c|c|c|c|c|c}
\hline
\rowcolor[HTML]{CCCCCC} 
\multicolumn{1}{c|}{\cellcolor[HTML]{CCCCCC}\textbf{Method}} & \multicolumn{2}{c|}{\cellcolor[HTML]{CCCCCC}\textbf{Accuracy}} & \multicolumn{2}{c|}{\cellcolor[HTML]{CCCCCC}\textbf{F1-Score}} & \multicolumn{2}{c|}{\cellcolor[HTML]{CCCCCC}\textbf{NMI}}      & \multicolumn{2}{c|}{\cellcolor[HTML]{CCCCCC}\textbf{ARI}}      & \multicolumn{2}{c|}{\cellcolor[HTML]{CCCCCC}\textbf{AMI}}      & \multicolumn{2}{c}{\cellcolor[HTML]{CCCCCC}\textbf{Time Corr.}} \\ \hline
Actor network features                                      & \multicolumn{2}{c|}{0.250}                                     & \multicolumn{2}{c|}{0.152}                                     & \multicolumn{2}{c|}{0.120}                                     & \multicolumn{2}{c|}{0.096}                                     & \multicolumn{2}{c|}{0.017}                                     & \multicolumn{2}{c}{0.118}                                       \\ \hline
\rowcolor[HTML]{F3F3F3} 
\multicolumn{13}{l}{\cellcolor[HTML]{F3F3F3}\textbf{1. Graph2Vec}}                                                                                                                                                                                                                                                                                                                                                                                            \\ \hline
Frequent clusters                                           & \multicolumn{2}{c|}{0.146}                                     & \multicolumn{2}{c|}{0.086}                                     & \multicolumn{2}{c|}{0.195}                                     & \multicolumn{2}{c|}{0.173}                                     & \multicolumn{2}{c|}{0.117}                                     & \multicolumn{2}{c}{0.263}                                       \\
Known actor name                                            & \multicolumn{2}{c|}{0.299}                                     & \multicolumn{2}{c|}{0.242}                                     & \multicolumn{2}{c|}{0.127}                                     & \multicolumn{2}{c|}{0.103}                                     & \multicolumn{2}{c|}{0.057}                                     & \multicolumn{2}{c}{-0.085}                                      \\ \hline
\rowcolor[HTML]{EFEFEF} 
\cellcolor[HTML]{F3F3F3}\textbf{2. Sequential}              & RW                                        & SPW               & RW                                        & SPW               & RW                            & SPW                           & RW                            & SPW                           & RW                            & SPW                           & RW                 & SPW                                        \\ \hline
All clusters                                                & 0.479                                     & 0.313             & 0.386                                     & 0.201             & 0.274                         & 0.281                         & 0.252                         & 0.227                         & 0.114                         & 0.068                         & 0.298              & 0.185                                      \\
Frequent clusters                                           & \cellcolor[HTML]{FFF2CC}0.771             & 0.681             & \cellcolor[HTML]{FFF2CC}0.665             & 0.526             & 0.333                         & 0.367                         & 0.315                         & 0.335                         & 0.160                         & 0.172                         & 0.459              & 0.635                                      \\
Known actor name                                            & 0.736                                     & 0.542             & 0.592                                     & 0.366             & 0.572                         & 0.332                         & 0.509                         & 0.297                         & 0.261                         & 0.126                         & 0.625              & \cellcolor[HTML]{FFF2CC}0.701              \\
Known actor type                                            & 0.764                                     & 0.688             & 0.646                                     & 0.552             & 0.425                         & \cellcolor[HTML]{FFF2CC}0.591 & 0.408                         & \cellcolor[HTML]{FFF2CC}0.544 & 0.180                         & \cellcolor[HTML]{FFF2CC}0.277 & 0.282              & 0.511                                      \\ \hline
\rowcolor[HTML]{EFEFEF} 
\cellcolor[HTML]{F3F3F3}\textbf{3. Temporal}                & RW                                        & SPW               & RW                                        & SPW               & RW                            & SPW                           & RW                            & SPW                           & RW                            & SPW                           & RW                 & SPW                                        \\ \hline
All clusters                                                & 0.389                                     & 0.222             & 0.288                                     & 0.131             & 0.298                         & 0.298                         & 0.278                         & 0.279                         & 0.130                         & 0.133                         & 0.253              & 0.193                                      \\
Frequent clusters                                           & 0.486                                     & 0.535             & 0.395                                     & 0.383             & 0.331                         & 0.326                         & 0.313                         & 0.307                         & 0.172                         & 0.168                         & 0.379              & \cellcolor[HTML]{FFF2CC}0.517              \\
Known actor name                                            & 0.583                                     & 0.521             & 0.450                                     & 0.350             & 0.453                         & 0.461                         & 0.402                         & 0.411                         & 0.209                         & 0.204                         & 0.453              & 0.491                                      \\
Known actor type                                            & \cellcolor[HTML]{FFF2CC}0.778             & 0.708             & \cellcolor[HTML]{FFF2CC}0.674             & 0.601             & \cellcolor[HTML]{FFF2CC}0.644 & 0.617                         & \cellcolor[HTML]{FFF2CC}0.618 & 0.580                         & \cellcolor[HTML]{FFF2CC}0.480 & 0.418                         & 0.282              & 0.356                                      \\ \hline
\end{tabular}
}
\end{table}

\autoref{tab:evaluation} reports the Accuracy and F1 measure of the classification models. Our walk-based embeddings provide higher accuracy and F1-score than the baseline models, and pruning strategies improve performances as the \textit{All clusters} approach consistently obtains the worst results. The best results are obtained using either \textit{Temporal Known Actor Type} or \textit{Sequential Frequent Clusters}. Overall, we found that \textit{Random walks} provides better results than \textit{Shortest Path Walks} in both sequential and temporal models.

\begin{figure}
\centerline{\includegraphics[width=\linewidth]{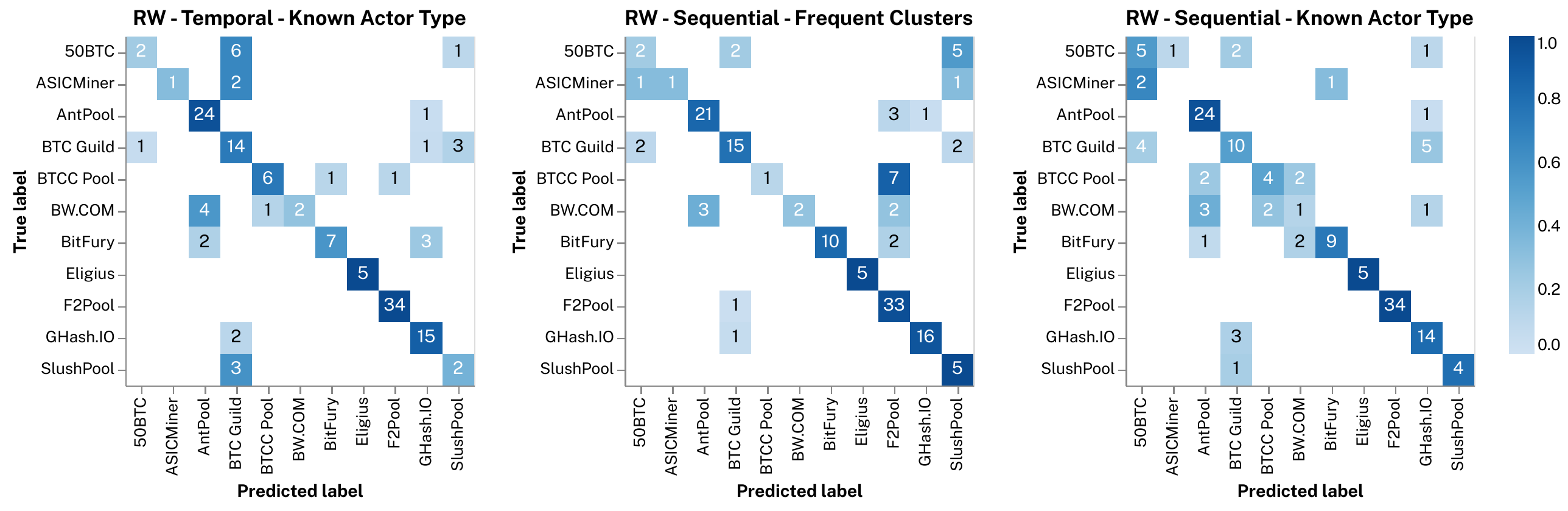}}
\caption{Confusion matrix of the top-3 classification models}
\label{fig:confusion}
\end{figure}

\autoref{fig:confusion} presents the confusion matrices of the top-3 classification models: \textit{Temporal Known Actor Type}, \textit{Sequential Frequent Clusters}, and \textit{Sequential Known Actor Type}. We observe that most of the actors for which we have many samples (e.g., AntPool, F2Pool, Ghash.IO) are well classified. Mining pools with few examples are more prone to errors, particularly ASICMiner, which has only three occurrences in our dataset and is mostly wrongly classified. This stresses the importance of having enough learning data in future works.
%\nat{In practice, the taint flow can be slow to extract depending on the $\rho_{min}$ and $time_{max}$. Nonetheless, the result shows that our embedding methods work well even with the simple model and limited taint flow data.}

\subsection{Actor Clustering Task}

We assess if source actors can be discovered with the unsupervised approach by training k-means clustering on flow embeddings and checking the the cluster founds with the known sources. We train k-means models from 2 to 11 clusters and select the model with the highest \textit{Silhouette} score. \autoref{tab:evaluation} reports the clustering evaluation with three standard scores: normalized mutual information (NMI), adjusted rand score (ARI), and adjusted mutual information (AMI).

We found that \textit{Known Actor Type} clusters are the most efficient method for sequential and temporal embeddings. In contrast to the classification task, the \textit{Shortest Path Walk} provides a better clustering for sequential embedding. Nonetheless, the three clustering metrics are relatively low. As source time and identity are highly correlated, we suspect the clustering might mix these two aspects and form the cluster containing pools in the same period.

%especially the AMI, which measures the mutual information adjusted against chance. 
%we suspect that the embeddings of mining pools are closer in the distance, thus forming fewer clusters than the number of mining pools. The result implies that it is not yet possible to use this technique to identify actors in a fully unsupervised way.

\subsection{Time Correlation}

%\begin{figure}
%\centerline{\includegraphics[width=\linewidth]{figures/tsne_shortest_3}}
%\caption{T-SNE projections of selected shortest path walk embeddings}
%\label{fig:tsne_shortest}
%\end{figure}

As we have observed in \autoref{fig:tsne}, the vectors of taint flow also embed a notion of time. This can be explained by Bitcoin's highly dynamic ecosystem in which actors appear, disappear, rise, and fall in popularity over time. Our approach depends on its source's identity and when this flow starts. Hence, it is possible that the embedding model group actors that frequently occur at the same period be closer in the embedding space.

To quantify how embeddings capture the evolution of taint flow patterns, we compute a \textit{time correlation} score by computing the Spearman correlation coefficient between the distance in time expressed in months and the distance in the embedding for all pairs of taint flows. The time correlation in \autoref{tab:evaluation} shows a high correlation for \textit{Sequential Known Actor Name} 
%and \textit{Sequential Frequent Clusters}. 
The correlation is lower in temporal embeddings compared with sequential ones.

Interestingly, the temporal aspect is poorly captured when using actor types instead of their identity. %, while the identity prediction task works well.
The correlation of \textit{Known Actor Type} suggests that despite the rise and fall in popularity of specific actors, the type of actors reached by a flow from a particular source tends to stay constant, as shown in \autoref{fig:tsne}. This result highlights the importance of using time-independent vocabulary, in this case, the role of the actors, to train the unbias embedding model.
%in which the \textit{Known Actor Type} projections show a stronger separation of some actors and a less linear shape of the embedding, which is characteristic of the temporal nature of sources.

%%%% REMY : I think the figure is not correct, the order of figures should be the same as the other one. We should keep only one of the two ?
% \begin{figure}
% \centerline{\includegraphics[width=\linewidth]{figures/tsne_random_6}}
% \caption{T-SNE projections of random walk embeddings}
% \label{fig:random}
% \end{figure}

% \subsection{Time Correlation of the Taint Flows}

\section{Discussion and Conclusion}
\label{conclusion}
In this work, we propose original methods to extract taint flows that take into account the temporal aspect of the Bitcoin transaction network and represent them using graph embedding techniques. We train classification and clustering models to evaluate how the tainted flows of coins can be used to identify the source actors in the Bitcoin transaction network.

Our experiments with mining pool taint flows show that although we could not reach a perfect precision in the actor identification task, a simple supervised approach yields a high accuracy. Unsupervised clustering is less convincing at this stage but could be improved by taking time into account and increasing the number of observations. The analysis of those results highlights what makes a taint flow characteristic of its source actor:
1) the starting time is significant as actors in the Bitcoin network emerge and disappear over time, and
2) the identity of encountered actors is not the only relevant element since we can also reach a good result using the characteristics of actors, especially their roles (actor type).
This stresses the importance of using labeled data to improve model performance and raises another research direction to infer actor roles from on-chain data.

Our work demonstrates the relevance of using taint flows to characterize their source. However, we can achieve a better model performance with more taint flow data and more sophisticated classification and clustering models. Additional information such as the country of origin, network centrality, and more precise actor types, could also be a direction of improvement.
%We stress that taint flow is another promising approach to classifying source actors that take into account the temporal aspect of the Bitcoin transaction network.
Our method could be applied to other cryptocurrencies or other forms of diffusion, such as information in social media, by appropriately adapting the diffusion flow's construction. In future work, we intend to apply this approach to characterize money flows in other domains, particularly illegal and cybercrime activities, as well as propose a new technique to extract and explain meaningful patterns from those flows.

% - Extracting the money flow is not new. But the pruning, graph model (temporal and pattern), and graph embedding should be the contribution of our work?
% - money flows in general is not new, but the way it's done (until dissolution) is new no ? and the pruning if pruning is keeping only some actors and summarizing in-between
% - and basically the bricks are not new, it's the way to assemble them which is new (like 90\% of research)

\section*{Acknowledgement}
This project was partly founded by BITUNAM grant ANR-18-CE23-0004.

\bibliographystyle{splncs03} % We choose the "plain" reference style
\bibliography{refs} % Entries are in the refs.bib file

\end{document}